# Single crystal $MgB_2$ with anisotropic superconducting properties

M. Xu[*†], H. Kitazawa[*], Y. Takano[*], J. Ye[*], K. Nishida[*], H. Abe[*], A. Matsushita[*] and G. Kido[*]

[*]*National Institute for Materials Science, 1-2-1, Sengen, Tsukuba 305-0047, Japan*
[†]*Japan Science and Technology Corporation, 2-1-6, Sengen, Tsukuba 305-0047, Japan*

**The discovery of superconductor in magnesium diboride $MgB_2$ with high $T_c$ (≈39 K) has raised some challenging issues; whether this new superconductor resembles a high temperature cuprate superconductor(HTS) or a low temperature metallic superconductor; which superconducting mechanism, a phonon- mediated BCS[1,2] or a hole superconducting mechanism[3,4] or other new exotic mechanism may account for this superconductivity; and how about its future for applications. In order to clarify the above questions, experiments using the single crystal sample are urgently required. Here we have first succeeded in obtaining the single crystal of this new $MgB_2$ superconductivity, and performed its electrical resistance and magnetization measurements. Their experiments show that the electronic and magnetic properties depend on the crystallographic direction. Our results indicate that the single crystal $MgB_2$ superconductor shows anisotropic superconducting properties and thus can provide scientific basis for the research of its superconducting mechanism and its applications.**

According to the phase diagram[5], it is preferable to grow single crystalline $MgB_2$ in a closed system for the high vapor pressure of Mg and high melting point of B. The single crystals $MgB_2$ used in this study were grown by the vapor transport method[6] in a sealed molybdenum crucible with the raw materials of Mg (99.99%) chunk and B (99.9%) chunk. Small thin plate single crystals with the sizes of $0.5\times0.5\times0.02$ mm$^3$ were selected from the crucible's inner surface. The crystal structural analysis was carried out by the X-ray precession camera with Mo target (no filter). The X-ray precession photograph of the crystal, shown in Figure 1, clearly reveals the hexagonal crystal structure with the lattice parameters a=0.3047(1) nm and c=0.3404(1) nm. The composition of the crystals was determined as $MgB_2$ by Electron Probe Microanalyzer(JEOL JXA-8900R).

The dc magnetic properties were measured with a SQUID magnetometer (MPMS-5S, Quantum Design) at applied field parallel to *c*-axis (*H//c*) and parallel to *ab*-plane (*H//ab*). Figure 2a shows temperature dependences of the zero-field-cooled (ZFC) and the field-cooled (FC) dc magnetization (M-T) curves of the single crystal sample in 1 mT field at *H//c* and *H//ab*. M-T measurements exhibit the same superconducting transition ($T_c^{onset}$=38.6 K) at *H//c* and *H//ab*. The magnetization hysteresis (M-H curve) was measured at several temperatures up to 5 T. Figure 2b shows the M-H curve at 5 K in the applied field up to 5 T at *H//c* and *H//ab*. The results of magnetization measurements, shown in Figure 2, clearly demonstrated the highly anisotropic magnetic character for this new superconductor.

The resistance of the sample as a function of temperature- and the magnetic field was measured with a Quantum Design PPMS-AG system using the standard four-probe ac method. The temperature dependence of the electrical resistance from 5 K to 280 K at 0 T, shown in Figure 3 with the onset $T_c$ value of 38.6 K, can be well fit by $R=R_0 + R_1T+R_2T^2$ with $R_0$=0.5449 mΩ, $R_1=7.8535\times10^{-3}$ mΩ/K and $R_2=5.0503\times10^{-5}$ mΩ/$K^2$ in the normal state. The insets in Figure 3 show the magnetic field dependence of the electrical resistance under magnetic fields up to 9 T at *H//c* and *H//ab*.

Here some superconducting parameters were determined by our experiments. From the M-H curves, the lower critical fields, $H_{c1}(T)$ for *H//c* and *H//ab* were determined and plotted as a function of the temperature for the single crystal as shown in Figure 4. Extrapolation of the plot gives the $H^{//c}{}_{c1}(0)$ and $H^{//ab}{}_{c1}(0)$ values of 27.2 mT and 38.4 mT, respectively. The upper critical fields, $H_{c2}(T)$ for *H//c* and *H//ab* at applied magnetic fields up to 9 T, were determined using the resistive onset temperature from the insets of Figure 3 and also plotted in the Figure 4. The $H_{c2}(T)$ curves for *H//c* and *H//ab* all show the linear behavior far from the Tc. Therefore, linear extrapolation gives the $H^{//c}{}_{c2}(0)$ and $H^{//ab}{}_{c2}(0)$ values of 9.2 T and 25.5 T respectively. Assuming the dirty limit of the type-II superconductor, which the $H_{c2}(0)$ value is given by the formula[7] of $\mu_0 H_{c2}(0) = 0.7\mu_0 T_c(-dH_{c2}/dT)|_{Tc}$, $H^{//c}{}_{c2}(0)$ and $H^{//ab}{}_{c2}(0)$ are found to be 7.7 T and 19.8 T respectively. Using the anisotropic Ginzburg-Landau (GL) formulas, $H^{//c}{}_{c2}=\phi_0/(2\pi\xi^2{}_{ab})$ and $H^{//ab}{}_{c2}=\phi_0/(2\pi\xi_{ab}\xi_c)$, the GL coherence length $\xi^{//c}(0)$ and $\xi^{//ab}(0)$ at zero temperature, can be estimated to be 2.5 nm and 6.5 nm respectively. These values of $\xi^{//c}(0)$ and $\xi^{//ab}(0)$ are larger than the typical values (1–2 nm) of high-temperature superconductors (HTS). The similar result was obtained from polycrystalline sample[8].

As a summary, we estimated some superconducting parameters from the single crystal $MgB_2$ superconductor and found that this new superconductor shows anisotropic superconducting properties with an anisotropy ratio $\gamma= H^{//ab}{}_{c2}(0)/H^{//c}{}_{c2}(0) \approx 2.6$, implying an anisotropy of the coherence length $\xi^{//ab}(0)/\xi^{//c}(0) \approx 2.6$ and a mass anisotropy rate $m_{ab}/ m_c \approx 0.15$. The value of anisotropy ratio γ is higher than the value of 1.7 for the aligned $MgB_2$ crystallites[9] and 1.8-2.0 for the c-axis oriented $MgB_2$ thin films[10]. Although the anisotropy in $MgB_2$, is much smaller than in the highly anisotropic high temperature cuprate superconductors(HTS) and graphite intercalation(GIC) superconductors[11], it nonetheless significantly affects the electronic and magnetic properties of $MgB_2$. Therefore, further investigation of the properties of $MgB_2$ will be very important for understanding its superconducting mechanism and applications.

Acknowledgements:

We thank Dr. N. Tsujii for technical guidance, Drs. M. Imai, Dr. H. Gu, Prof. Z. Jiao and G. Cao for useful discussions. This work was supported by the Japan Science and Technology Corporation (JST).

**Correspondence and requests for materials should be addressed to M.X. (e-mail: XU.Mingxiang@nims.go.jp).**

Figure legends:

Figure 1 Zero layer X-ray precession photograph of the crystal in [0 0 1] zone axis.

Figure 2 Magnetic properties of superconductivity (corrected by the demagnetization effect). **a**, Magnetization of the single crystal as a function of temperature after cooling in zero field and cooling in a field of 1 mT at *H//c* and *H//ab*. Inset shows the enlarger between 34 and 40 K, showing the same superconducting transition ($T_c^{onset}$=38.6 K) at *H//c* and *H//ab*. **b**, Magnetization of the single crystal as a function of applied field up to 5 T at 5 K for *H//c* and *H//ab*.

Figure 3 Electrical resistance of the single crystal as a function of temperature at 0 T. The upper and lower insets show the electrical resistance as a function of temperature and magnetic fields up to 9 T at *H//c* and *H//ab* respectively.

Figure 4 Magnetic field-temperature phase diagram for $MgB_2$ single crystal obtained from electrical and magnetic experiments. The lower and upper critical fields, $H_{c1}(T)$ and $H_{c2}(T)$ as a function of temperature at *H//c* and *H//ab*. The values of $H_{c1}(T)$ were defined as the magnetic field where the initial slope of $M_{up}$ curve meets the extrapolation curve of $(M_{up}+M_{down})/2$. The values of $H_{c2}(T)$ were determined using the resistive onset temperature from the insets of figure 3.

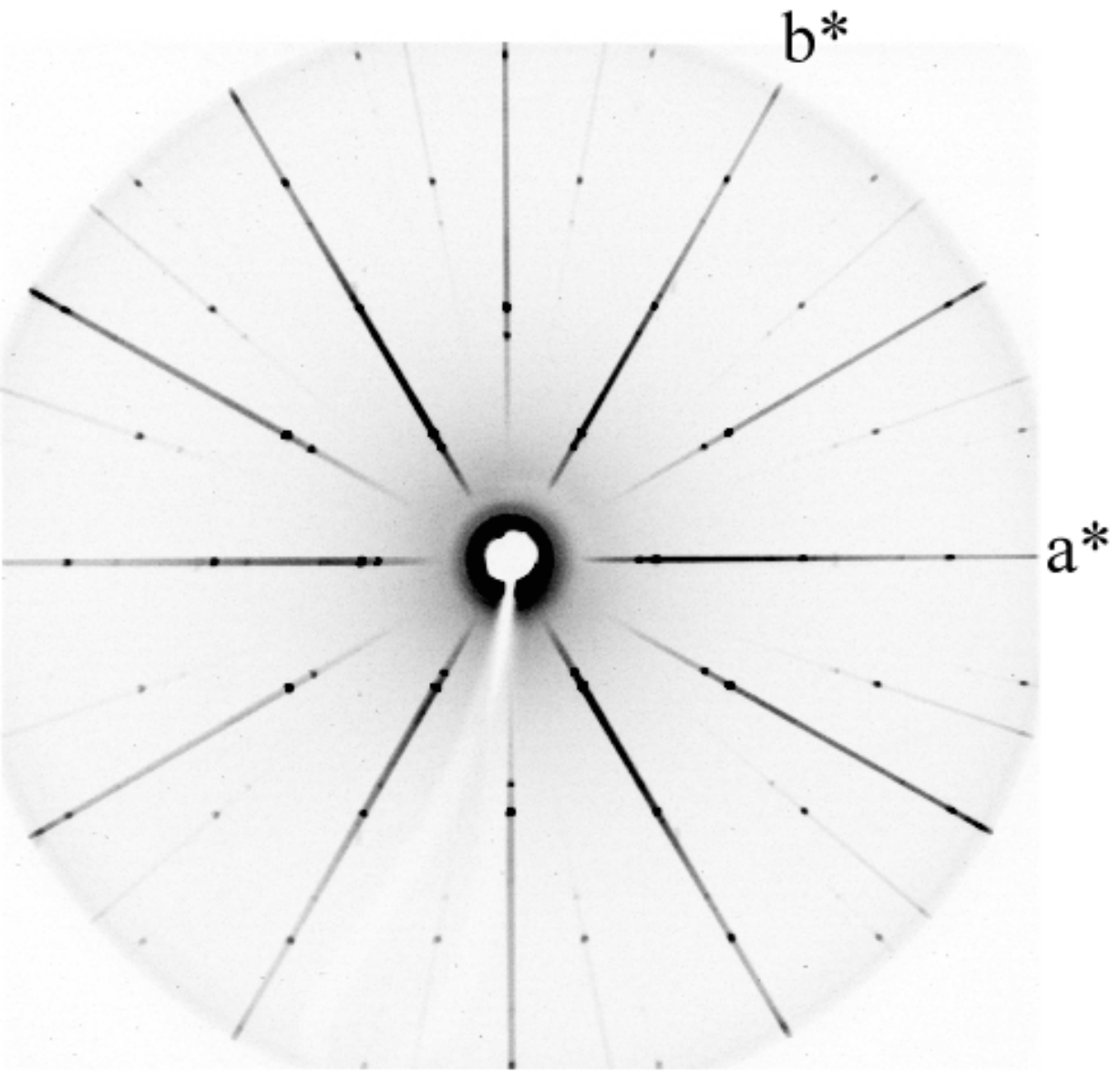
b*
a*

**Fig. 1**

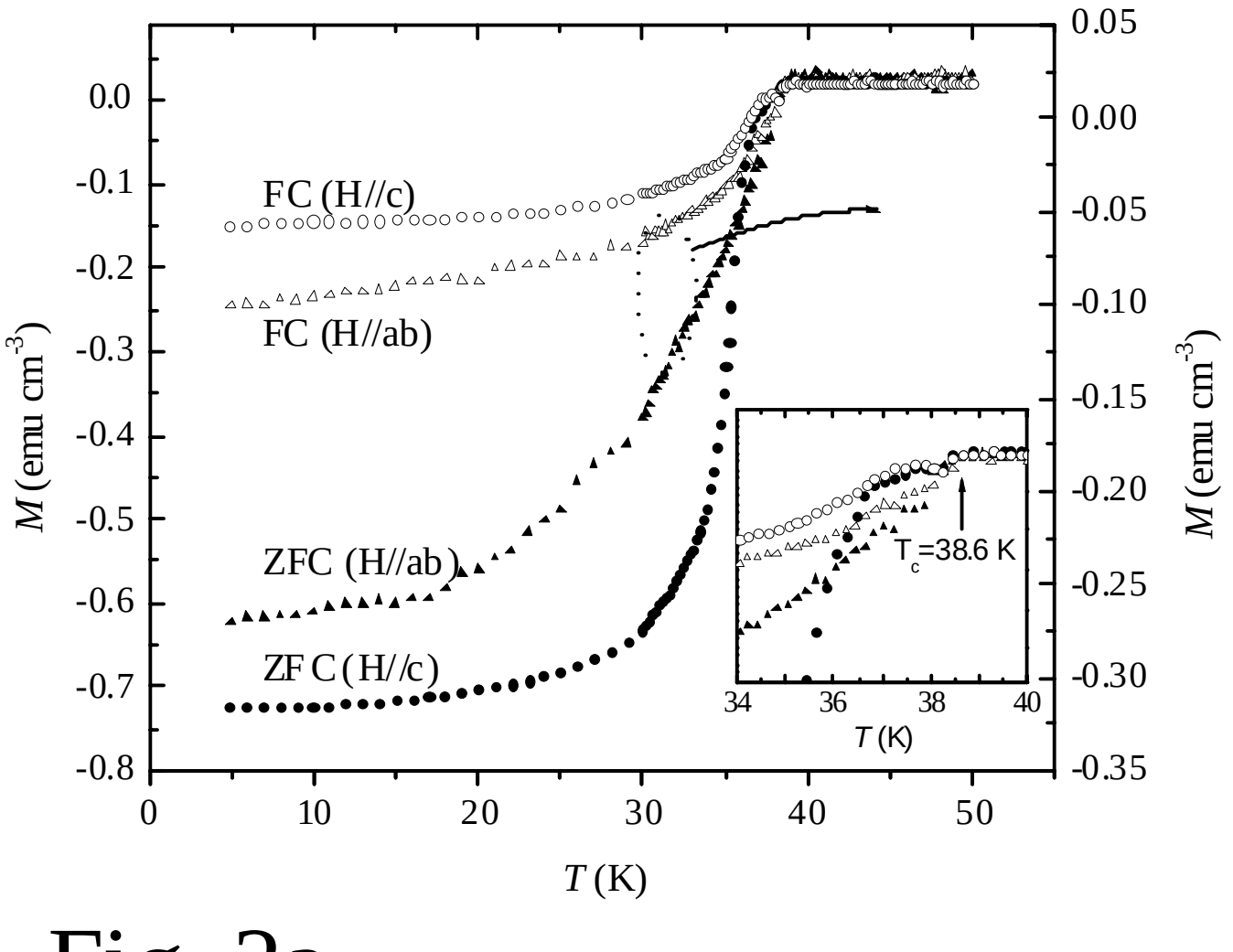

Fig. 2a

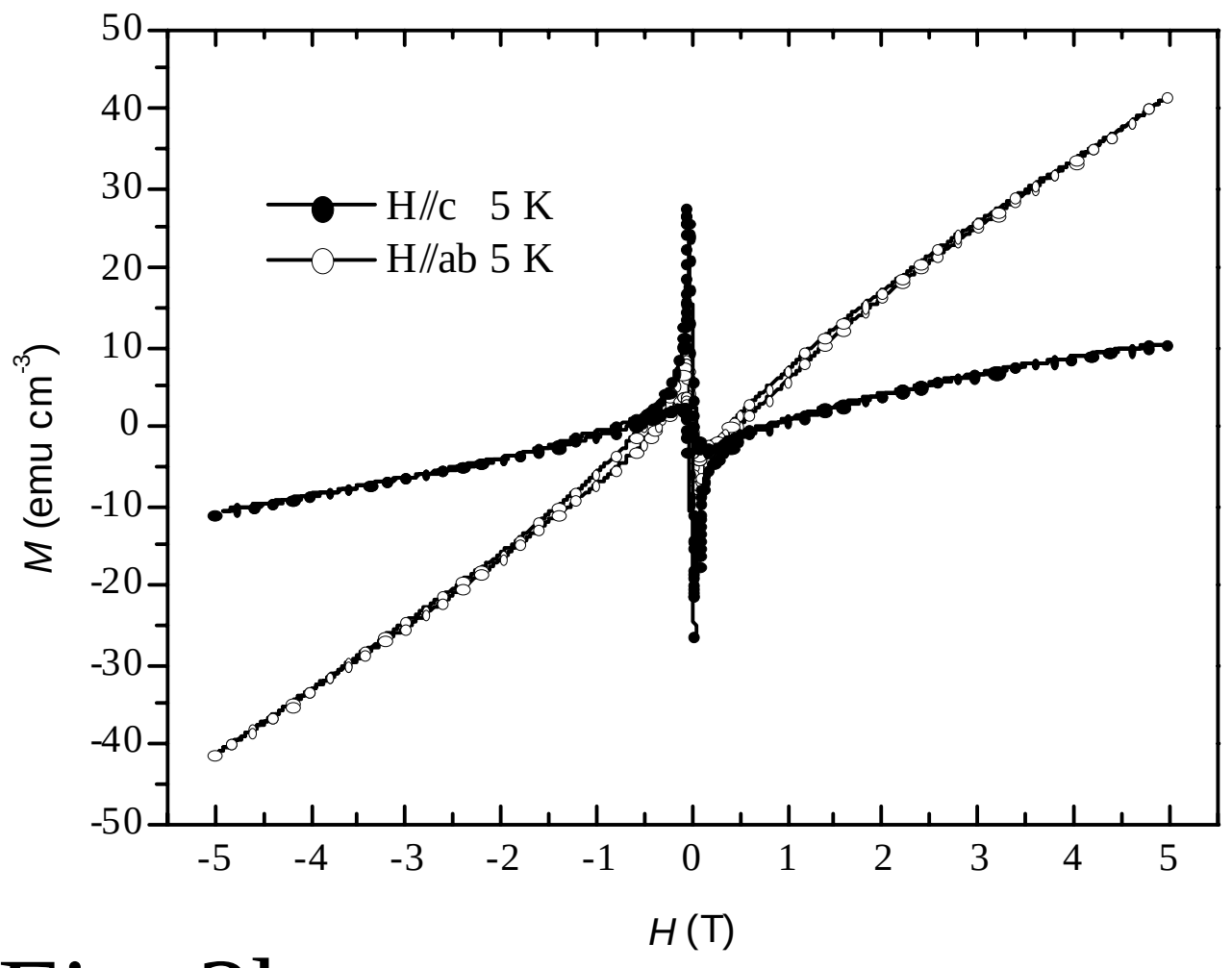

Fig. 2b

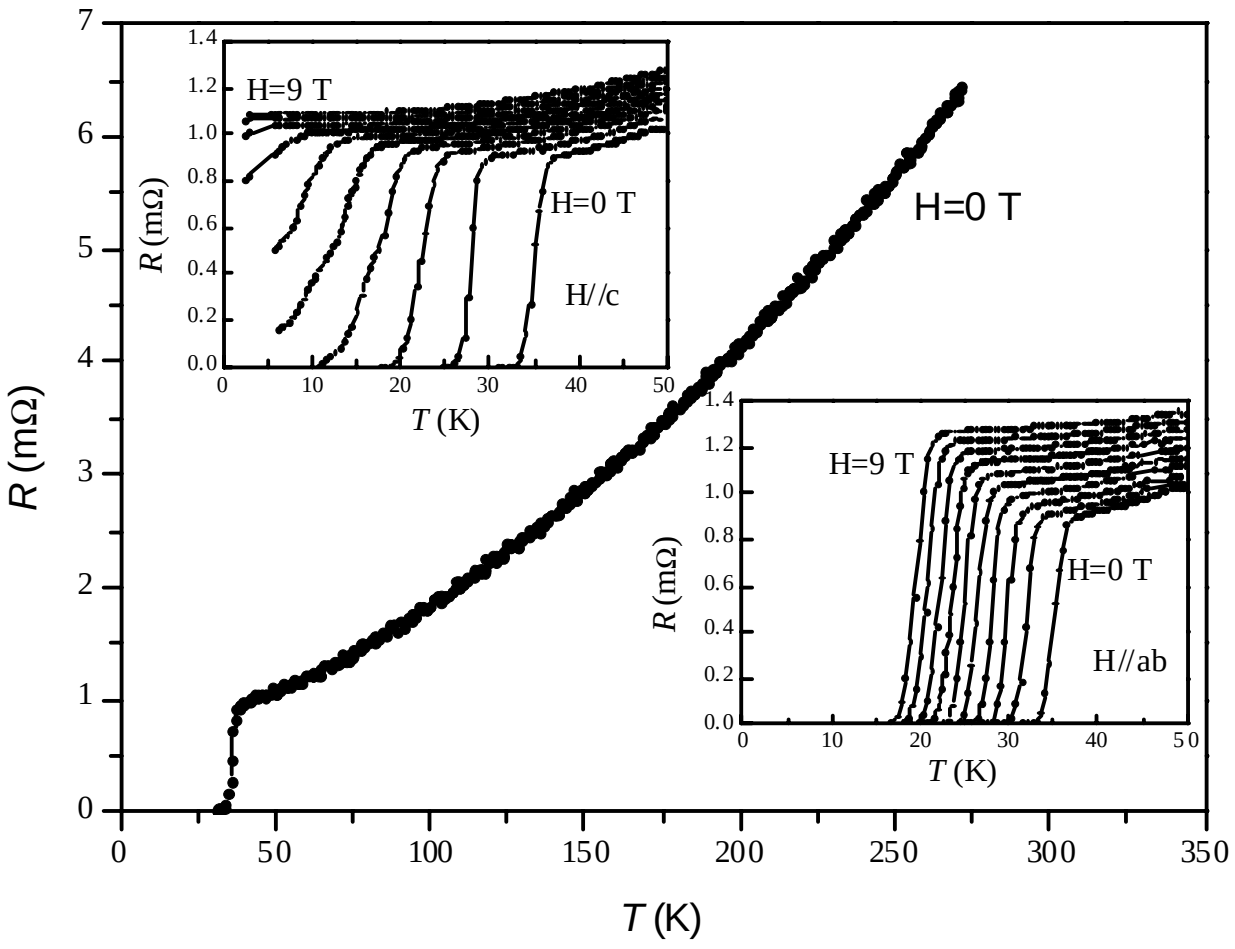

Fig. 3

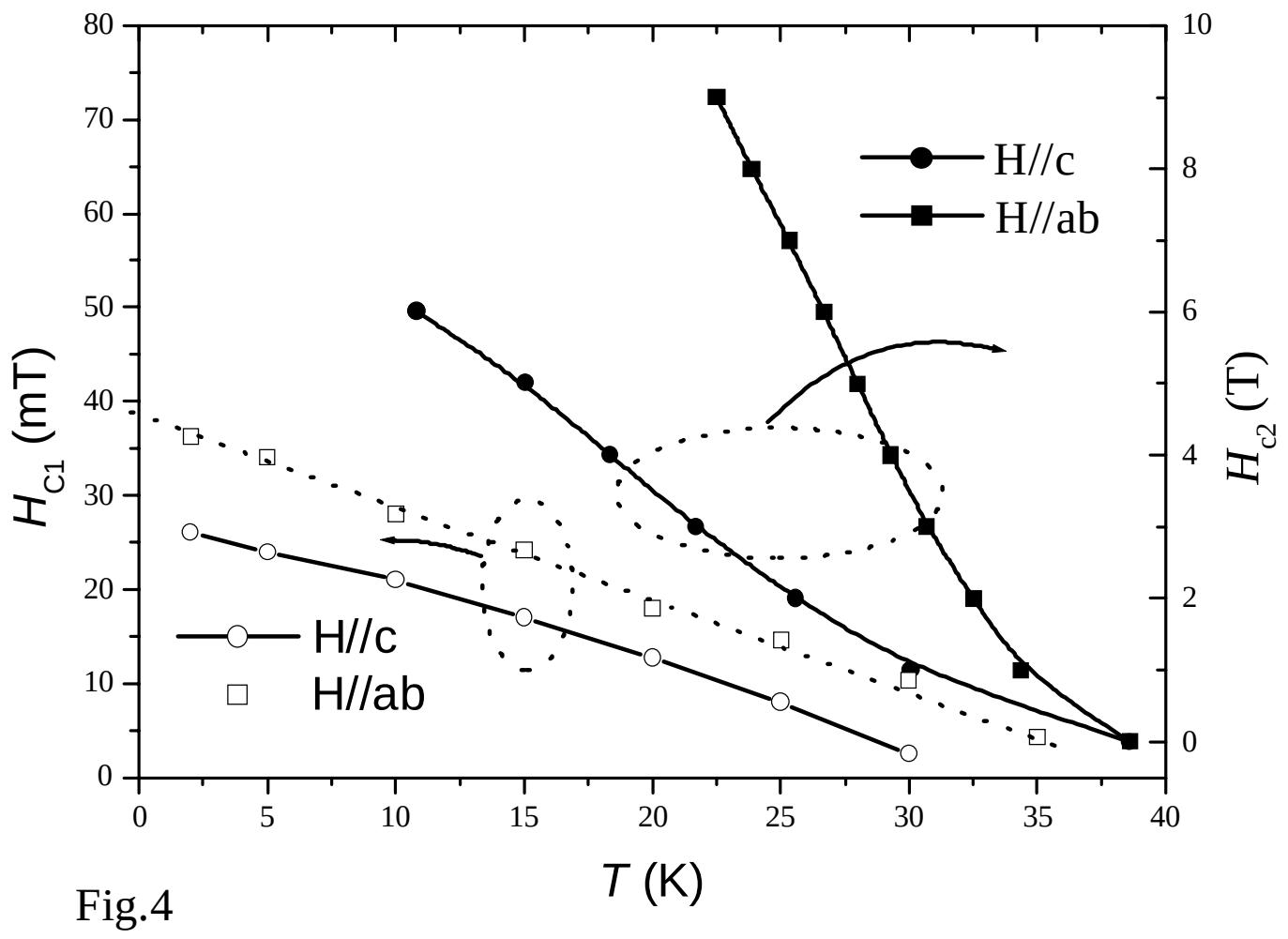

Fig.4